\documentstyle[11pt,paspconf,epsfig,graphics,color]{article}

\begin{document}

\title{Keck Spectroscopy of Giant Arcs in Abell
2390\altaffilmark{1}}

\author{B. L. Frye, T. J. Broadhurst, H. Spinrad, and A. Bunker}
\affil{Astronomy Department, University of California,
    Berkeley, CA 94720}


\altaffiltext{1}{Based on observations obtained at the W.M.  Keck Observatory,
Hawaii, which is operated jointly by the California Institute of Technology and
the University of California.}



\begin{abstract}

 We present new Keck observations of giant arcs in the cluster
Abell 2390.  High resolution two-dimensional spectra of two arcs show
metal lines at $z=4.040\pm 0.005$ with Ly$\alpha$ emission spatially
separated from the stellar continuum.  In addition, spectroscopy along
the notorious `straight' arc reveals two unrelated galaxies, at
z=0.913 and z=1.033, with absorption lines of MgII and FeII at z=0.913 seen
against the more distant object, indicating the presence of a large
gaseous halo.  
\end{abstract}




\section{The Z=4.04 Arcs}

  New high resolution spectra are presented for the high-redshift
$z=4.04$ lensed system behind A2390. Figure 1 shows a section of the
spectrum aligned along the long axes of the two main arcs.  This
clearly shows that the Ly$\alpha$ emission is spatially-separated from
the continuum light and redshifted with respect to the interstellar
lines, indicating an outward flow of enriched gas thought to be
typical of starburst galaxies (Lequeux et al. 1995). All stellar and
interstellar features are common to both spectra, confirming these
arcs are images of one single highly-magnified galaxy. Note the
Ly$\alpha$ absorption is seen only in the southern portions of
both arcs, coincident with the stellar continuum and with no associated
Ly$\alpha$ emission, indicating absorption of this line 
where the HI column is high. A lens model for this system is
discussed in Frye \& Broadhurst (1998). Keck infrared observations
were also taken to study the stellar populations of this galaxy, by
comparing flux ratios on opposite sides of the 4000 \AA \ break
(Bunker, et al. these proceedings).

\section{The Straight Arc}

  The famous `straight' arc was first observed spectroscopically by
Pell\'o, et al. 1991 at z=0.913. The straightness and length of this
arc led to difficulties in finding a reasonable lens model (Kassiola,
et al. 1992).  We present here new Keck spectra taken along this arc
confirming the known redshift, z=0.913, but also revealing that the 
redshift of component A of this arc (in the notation of Pell\'o et al) 
is in fact an unrelated galaxy at z=1.033.  Thus
the straight arc comprises two unrelated galaxies, close in projection.
Interestingly, low
ionization metal lines are found at $z=0.913$ against the spectrum of the 
higher redshift galaxy, indicating an extended gaseous halo around
this lower redshift galaxy, similar
to the halos detected against bright QSOs (Bergeron, et al. 1992, 
Steidel, et al. 1994, Churchill, et al. 1996) as shown in  Figure 3.

This progress report on giant arcs in A2390 is part of a larger
Keck arclet redshift survey to carry out multi-slit spectroscopy of
galaxies behind massive clusters.  

%



\acknowledgments We thank Art Wolfe, Genevieve Soucail, Richard
Ellis, and Rosa Pell\'o for useful conversations.


\newpage
\centerline {\bf{Figure Captions}}
\medskip
\bigskip

Figure 1.  Two-dimensional spectrum of two arcs at z=4.040, taken
along the long axes of the arcs.  The northern and southern components extend 5
and 3 arcsec above and below the central cluster elliptical, the bright white
central strip.  The blue end is at the left hand side of the chip and the pixel
scale is 0.21 arcsec/pix.  Ly$\alpha$ emission, seen in each component as a
bright spot in the blue is spatially separated from the broad clearly visible
continuum and the damped Ly$\alpha$ absorption.

\bigskip

Figure 2.  High resolution Keck spectra of two arcs at z=4.040 (8 \AA \
resolution).  The upper two panels show the spectra of the southern
and northern arcs. (see Fig. 1).  The interstellar lines of
SiII, OI, CII, and SiIV are clearly seen in both images, used in
determining the redshift. These metal lines lie blueward of the centroid
of the Ly$\alpha$ emission by 300km/s. This behavior is common to
all high quality high redshift and local starburst galaxy spectra 
and indicates global outflow of gas.

\bigskip

Figure 3.  Keck Spectra taken at two different positions along the
Straight Arc.  Two unrelated galaxies are found at z=0.913 and
z=1.033, corresponding to components C and A respectively (in the
notation of Pell\'o, et al. 1991).  In addition interstellar lines of
MgII and FeII at the redshift of component C are seen in absorption
towards the higher redshift component A (indicated by dashed lines
extending downward into the middle panel), evidence for an extended
gaseous halo.

%


\begin{references}

\reference{}Bergeron, J., Christiani, S., \& Shaver, P. A. 1992, A\&A, 257, 417

\reference{}Bezecourt, J. \& Soucail, G. 1997, A\&A, 317, 661

\reference{}Bunker, A. J., Moustakas, L. A., Davis, M., Frye, B. L.,
Broadhurst, T. J., \& Spinrad, H. 1997, this volume, and astro-ph/9712173 

\reference{}Churchill, C. W., Steidel, C., C., \& Vogt, S. S. 1996, ApJ,
471,164

\reference{}Frye, B. L. \& Broadhurst, T. J. 1998, accepted ApJL, 
astro-ph/9712111

\reference{}Kassiola, A. \& Kovner, I. 1992, ApJL, 338, 33

\reference{}Lequeux, F., Kunth, D., Mas-Hesse, J. M., \& Sargent, W. L. W. 1995, A\&A, 301, 18L

\reference{}Pell\'o, R., Sanahuja, B., Le Borne, J.-F., Soucail, G., \&
                  Mellier, Y. 1991, ApJ, 366, 405

\reference{}Steidel, C. C., Dickinson, M, \& Persson, S. E. 1994, ApJL, 437, 75

\end{references}
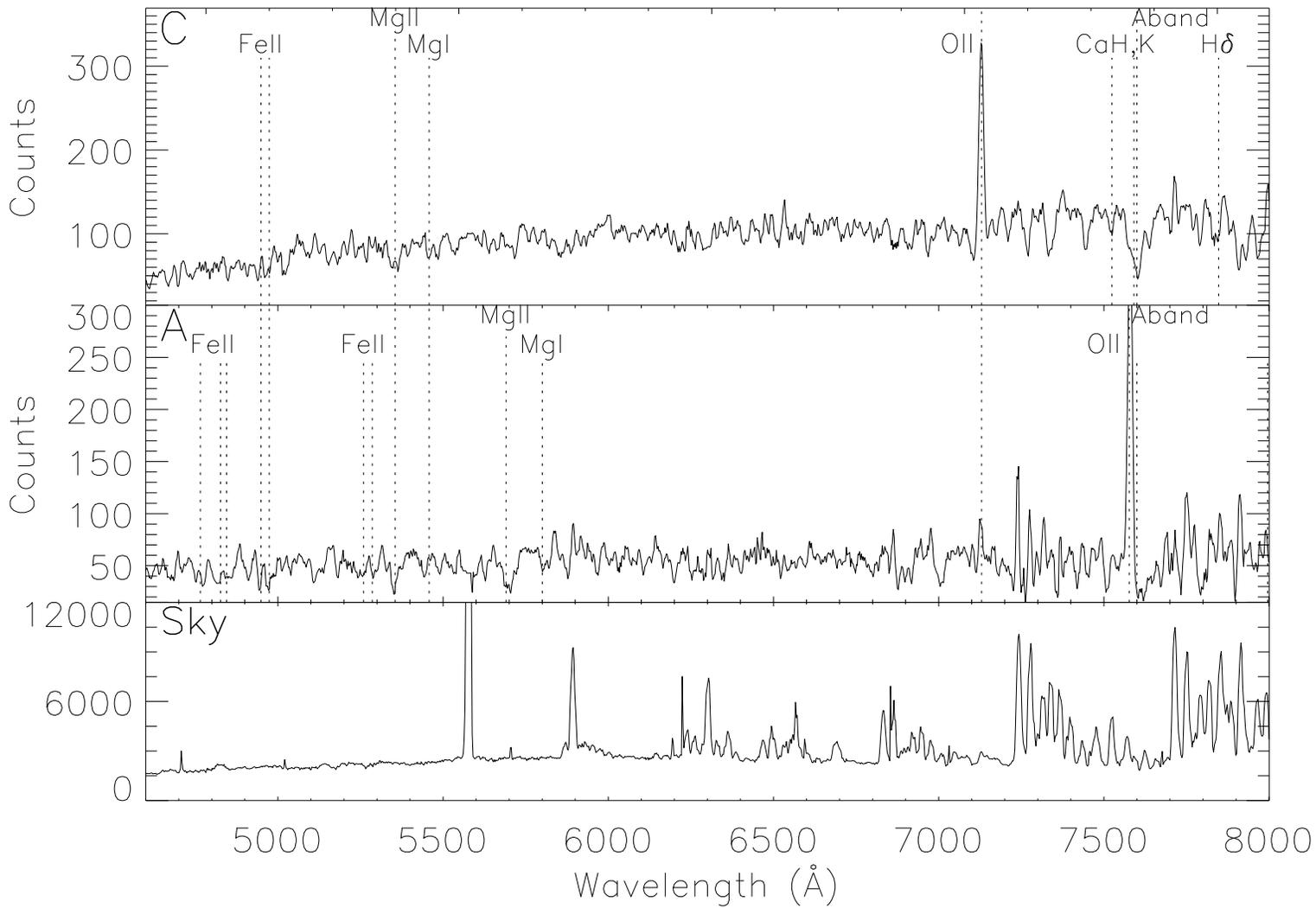
\end{document}